\documentclass[sigconf]{acmart}

\AtBeginDocument{%
  }

\setcopyright{acmlicensed}
\copyrightyear{2018}
\acmYear{2018}
\acmDOI{XXXXXXX.XXXXXXX}

\acmConference[Conference acronym 'XX]{Make sure to enter the correct
  conference title from your rights confirmation emai}{June 03--05,
  2018}{Woodstock, NY}
\acmISBN{978-1-4503-XXXX-X/18/06}

\usepackage{algorithm,algpseudocode,bbm}
\newcommand{\M}{\mathcal{M}}
\newcommand{\A}{\mathcal{A}}
\newcommand{\D}{\mathcal{D}}
\newcommand{\U}{\mathcal{U}}
\newcommand{\V}{\mathcal{V}}
\newcommand{\pa}{\mathtt{Pa}}

\theoremstyle{remark}
\newtheorem{remark}{Remark}
\usepackage{algpseudocode}
\begin{document}

\title{Multi Agent Influence Diagrams for DeFi Governance}



\author{Abhimanyu Nag}
\affiliation{%
    \institution{University of Alberta}
  \institution{Chainrisk Labs}
  \city{}
  \country{Canada}}

\author{Samrat Gupta}
\affiliation{%
  \institution{Chainrisk Labs}
  \city{}
  \country{India}}

\author{Sudipan Sinha}
\affiliation{%
  \institution{Chainrisk Labs}
  \city{}
  \country{UAE}}

\author{Arka Dutta}
\affiliation{%
  \institution{Chainrisk Labs}
  \city{}
  \country{UAE}}

\renewcommand{\shortauthors}{Nag, Gupta, Sinha, and Dutta}
\begin{abstract}
Decentralized Finance (DeFi) governance models have become increasingly complex due to the involvement of numerous independent agents, each with their own incentives and strategies. To effectively analyze these systems, we propose using Multi Agent Influence Diagrams (MAIDs)~\cite{Koller2001} as a powerful tool for modeling and studying the strategic interactions within DeFi governance. MAIDs allow for a comprehensive representation of the decision-making processes of various agents, capturing the influence of their actions on one another and on the overall governance outcomes. In this paper, we study a simple governance game that approximates real governance protocols and compute the Nash equilibria using MAIDs. We further outline the structure of a MAID in MakerDAO.
\end{abstract}

\maketitle

\section{Introduction}
Decentralized Finance (DeFi) governance protocols empower token holders to actively participate in shaping the future of the platform by voting on a variety of important decisions. These decisions can range from technical adjustments, such as parameter updates, to broader protocol improvements that impact the entire ecosystem. The process of deciding how to vote on a particular proposal is complex and influenced by multiple factors.
Firstly, voter behavior is often shaped by the actions and decisions of other voters. This phenomenon, known as social influence or herding behavior, can significantly impact the outcome of votes as individuals may be swayed by the perceived consensus or the actions of influential stakeholders within the network. Secondly, external factors, such as market sentiment, play a critical role in shaping voter decisions. Market trends, news events, and the overall economic climate can affect how token holders perceive the implications of a proposal and thus influence their voting behavior.
Understanding the dynamics of these voting games is crucial for ensuring the robustness of the system. For the analysis of such games, we propose to use Multi Agent Influence Diagrams (MAIDs). 
MAIDs combine the ideas of Bayesian Networks (BNs)~\cite{pearl2014probabilistic} and Influence Diagrams~\cite{howard2005influence} to describe decision problems related to multiple agents. MAIDs are graphical models used to represent decision-making scenarios involving multiple agents, each with their own objectives, beliefs, and actions. They extend the traditional influence diagram framework to accommodate the complexities of interactions and dependencies among multiple decision-makers.
In a MAID, nodes represent variables such as decisions, uncertainties, utilities, and influences, while directed edges indicate causal relationships or dependencies between them. Each agent is associated with a subset of nodes, representing their local decisions, beliefs, and preferences (see~\cite{Koller2001}).
MAIDs allow for the explicit representation of interactions between agents, including cooperation, competition, coordination, and negotiation. This enables the analysis of strategic interactions and the prediction of outcomes resulting from the decisions of multiple agents. There has been a lot of related work done by Google Deepmind (see~\cite{everitt2021agent,everitt2019understanding}) on agent incentives for artificial intelligence architectures which forms the underlying motivator to try and bring the same ideas to DeFi.
Just as Bayesian networks make explicit the dependencies between probabilistic variables, MAIDs make explicit the dependencies between the probabilistic variables that determine the decision taken by each agent in the game. Note that for every MAID, there is a corresponding extensive form game.

The paper is structured as follows. The next section introduces MAIDs. Section~\ref{model} describes a simple governance model. In Section~\ref{analysis}, we solve the MAID for Nash equilibria in a single-agent and in a multi-agent setting. Section~\ref{sec:maker} presents a MAID for MakerDAO governance. Finally, Section~\ref{sec:conclusion} concludes.



\section{Multi Agent Influence Diagrams (MAIDs)} \label{MAIDs}

In this section, we recall the necessary notions from Koller~and~Milch~\cite{Koller2001} to introduce MAIDs and the algorithm to compute Nash equilibria of MAIDs.

\begin{definition}[Chance, Decision, and Utility Variables]
We define the variables as follows:
    \begin{itemize}
        \item A \textit{chance variable} represents uncertain events or outcomes, similar to nodes in \textit{Bayesian networks}~\cite{pearl2014probabilistic} that affect each agent's decisions. The set of all chance variables is denoted by $\chi$.
        \item A \textit{decision variable} is a variable whose value is chosen by an agent and these are the decisions that each agent $ a \in \A$ can make. The set of all chance variables is denoted by $\D$.
        \item A \textit{utility variable} specifies a utility function for each agent $a \in \A$. The set of all utility variables is denoted by $\U$.
    \end{itemize}
We define $\V:= \chi \cup \D \cup \U$ the set of all variables.
\end{definition}
\begin{definition}[MAID]
    A \textit{MAID} is tuple $\M = \langle \A, \chi, \D, \U \rangle$, where $\A$ is the (finite) set of agents, $\chi$ the (finite) set of chance variables, $\D$ the (finite) set of decision variables, and $\U$ the (finite) set of utility variables. For each agent $a \in \A$ the individual sets of variables are $\langle \D_a, \U_a \rangle$ and thus $\D = \bigcup_{a \in \A} \D_a, \U = \bigcup_{a \in \A} \U_a$. Note that $\M$ defines a directed acyclic graph (DAG)~\cite{williams2018directed} structure where variables are nodes.
\end{definition}
\begin{definition}[Causal Influence Diagram]
    A \textit{Causal influence diagram (CID)}~\cite{everitt2021agent} for an agent $a \in \A$ is a tuple $(\V_a,E_a)$ where $(\V_a,E_a)$ is a directed acyclic graph (DAG) with a set of vertices $\V_a = \chi \cup \D_a \cup \U_a$ connected by directed edges $E_a \subseteq\V_a \times\V_a$.
\end{definition}

\begin{definition}[Parent Set]
    Let $\M = \langle \A, \chi, \D, \U \rangle$ be a MAID. For every variable $X \in \V$, $\pa(X)\subset \V\setminus \U$ is the set of all \textit{parent} nodes. By definition (and in keeping with the structure of a DAG) utility variables do not have child nodes. For any agent $a \in \A$, the parent set $\pa(D)$ for $D \in \D_a$, is the set of variables whose values agent $a$ knows when he chooses a value for D. 
\end{definition}
\begin{definition}[Conditional Probability Distribution (CPD)]
    Let $\M$ be a MAID. Then, for each chance variable $X \in \chi$, $\M$ induces a \textit{conditional probability distribution (CPD)} $\mathbb{P}(X| \mathbf{pa})$ for each instantiation (i.e for each instance of variable) $\mathbf{pa}\in\pa(X)$. Similarly, for each utility variable $U \in \U$, $\M$ induces a CPD $\mathbb{P}(U| \mathbf{pa})$ for each instantiation $\mathbf{pa}\in\pa(X)$.\footnote{there are more things on the utility. maybe we need to add them.}
\end{definition}
\begin{definition}[Decision Rule, Strategy]
A \textit{decision rule} for a decision variable $D \in \D$ is a function that maps each instantiation $\mathbf{pa}$ of parents $\pa(D)$ to a probability distribution over the domain $\mathtt{dom}(D)$. An assignment of decision rules to every decision $D \in \D_a$ for a particular agent in $a \in \A$ is called a \textit{strategy}.
\end{definition}
\begin{definition}[(Partial) Strategy Profile]
    An assignment $\sigma$ of decision rules to every decision $D\in \D$ is called a \textit{strategy profile}. A partial strategy profile $\sigma_\mathcal{E}$ is an assignment of decision rules to a subset $\mathcal{E} \in \D$.
\end{definition}
\begin{remark}
    Given a MAID $\M$, then a partial strategy profile $\sigma_\mathcal{E}$ induces a new MAID $\M{[\sigma_\mathcal{E}]}$ where the elements of $\mathcal{E}\subset \D$ are the chance variables and for each $D \in \mathcal{E}$, $\sigma_\mathcal{E}(D)$ is a CPD.
\end{remark}

\begin{definition}[Joint Distribution]
    If $\M$ is a MAID and $\sigma$ is a strategy profile for $\M$, then the joint distribution for $\M$ induced by $\sigma$, denoted $P_{\M[\sigma]}$, is the joint distribution over $\V$ defined by the Bayes network where:
\begin{itemize}
    \item the set of variables is $\V$;
    \item for $X,Y \in \V$, there is an edge $X \rightarrow Y$ iff $X \in \pa(Y)$;
    \item for all $X \in \V\setminus \D$, the CPD for $X$ is $\mathbb{P}(X)$;
    \item for all $D\in \D$, the CPD for $D$ is $\sigma(D)$.
\end{itemize}
\end{definition}
We are ready to motivate the definition of agent utility.
\begin{definition}[Expected Utility]
The expected utility that any agent $a \in \A$ anticipates in a MAID $\M$ when the agents play strategy profile $\sigma$ is

\begin{align}
     EU_a (\sigma) &=  \sum_{(u_1,u_2,\ldots,u_m) \in dom(\U_a)} P_{\M[\sigma]}(u_1,u_2, \ldots, u_m) \sum_{i = 1}^{m} u_i \\
    &=\sum_{U \in \U_a} \sum_{u \in dom(U)}  P_{\M[\sigma]}(U = u) \cdot u
\end{align}

\end{definition}


\begin{definition}[Optimal Strategy]
Let $\mathcal{E}$ be a subset of decision variables $\D_a$, and let $\sigma$ be a strategy profile. We say that $\sigma^{*}_\mathcal{E}$ is \textit{optimal for the strategy profile} $\sigma$ if, in the induced MAID $\M[\sigma_{\mathcal{E}}]$, where the only remaining decisions are those in $\mathcal{E}$, the strategy $\sigma_\mathcal{E}$ is optimal.

Formally, this means that for all strategies $\sigma'_{\mathcal{E}}$:

\begin{align*}
EU_a(\sigma_{-\mathcal{E}}, \sigma^{*}_\mathcal{E}) \ge EU_a(\sigma_{-\mathcal{E}}, \sigma'_\mathcal{E}).
\end{align*}
\end{definition}

\begin{definition}[Nash Equilibrium for a MAID]
A strategy profile $\sigma$ is a Nash equilibrium for a MAID $\M$ if for all agents $a \in \A$, $\sigma_{\D_a}$ is optimal for the strategy profile $\sigma$.
\end{definition}

Now that we understand the Nash equilibrium in a MAID, we present an algorithm that helps us compute the Nash Equilibria~\cite{nash1950non}.

\begin{proposition}[MAID and Extensive-Form Game~\cite{Koller2001}]
    Let $\M$ be a MAID. Then, there is a corresponding extensive-form game tree $\mathcal{T}$. For any strategy profile $\sigma$, the payoff vector for $\sigma$ in $\M$ is the same as for $\sigma$ in $\mathcal{T}$.
\end{proposition}
\subsection{Algorithm for Computing the Nash Equilibrium}
In order to compute the Nash equilibrium for a particular MAID $\M$, we will require the use of a few more definitions as follows: 

\begin{definition}[s-reachable]
A node $D'$ is \textit{$s$-reachable} from a node $D$ in a MAID $\M$ if there exists some utility node $U \in \U_D$ such that adding a new parent $D''$ to $D'$ would create an active path in $\M$ from $D''$ to $U$, given the evidence set $\pa(D) \cup \{D\}$.

Here, a path is considered active in a MAID if it is active in the same graph viewed as a Bayesian network (BN)~\cite{pearl2014probabilistic}.
\end{definition}

\begin{definition}[Relevance Graph]
The \textit{relevance graph} for a MAID $\M$ is a directed graph whose nodes are the decision nodes of $\M$. There is an edge directed from node $D$ to node $D'$ (denoted $D \rightarrow D'$) if and only if $D'$ is $s$-reachable from $D$ in $\M$.
\end{definition}

\begin{definition}[Strongly Connected Component]
A set $\mathcal{S}$ of nodes in a directed graph is a \textit{strongly connected component (SCC)} if for every pair of nodes $D\neq D' \in \mathcal{S}$, there exists a directed path from $D$ to $D'$. A \textit{maximal SCC} is an SCC that is not a strict subset of any other SCC.

We can find the maximal SCCs of a relevance graph in linear time, by constructing a \textit{component graph}~\cite{cormen2022introduction} whose nodes are the maximal SCCs of the graph. There is an edge from component $\mathcal{C}_i$ to component $\mathcal{C}_j$ in the component graph if and only if there is an edge in the relevance graph from some element of $\mathcal{C}_i$ to some element of $\mathcal{C}_j$.

The component graph is always acyclic, so we can define an ordering $C_1, C_2, \cdots C_m$ over the SCCs, such that whenever $i<j$  no element of $C_i$ is $s$-reachable from any element of $C_j$.
\end{definition}

\begin{definition}[Topological Ordering~\cite{cormen2022introduction}]
        Let $(\V, E)$ be a DAG. A topological ordering of $(\V, E)$ is a linear ordering of its vertices such that for every directed edge $(u, v)$ in $E$, vertex $u$ comes before $v$ in the ordering.
\end{definition}

Next, we present the algorithm for computing the Nash equilibria for a MAID.
\begin{algorithm}
\caption{Computing Nash Equilibria~\cite{Koller2001}}
\label{alg:maid}
\begin{algorithmic}
\Require Given a MAID $\M$
\Require a topological ordering $C_1, C_2, \cdots C_m$  of the component graph derived from the relevance graph for $\M$ 
\State Let $\sigma^{0}$ be an arbitrary fully mixed strategy profile
\For{$i = 0$ through $m-1$:}
        \State Let $\tau$ be a partial strategy profile for $C_{m-i}$ that is a Nash equilibrium in $\M[{\sigma^{i}}_{-C_{m-i}}]$
        \State Let $\sigma^{i+1} = ({\sigma^{i}}_{-C_{m-i}},\tau)$ 
\EndFor
\State Output $\sigma^{m}$ as an equilibrium of $\M$
\end{algorithmic}
\end{algorithm}

The algorithm processes each Strongly Connected Component (SCC) in reverse order, determining an equilibrium strategy profile for each SCC in the Multi-Agent Influence Diagram (MAID). This is done based on the previously selected decision rules, with arbitrary decision rules assigned to decisions that are not pertinent to the current SCC. In this derived MAID, the only decision nodes that remain are those within the current SCC, while all other decision nodes are treated as chance nodes. To find the equilibrium in this transformed graph, a subroutine designed to identify equilibria in games is used. The game is then transformed into a game tree, and a standard game-solving algorithm is applied to identify a decision rule that maximizes the expected utility for the individual agent. 

Before stating the result of correctness of the algorithm, we need the following definition.
\begin{definition}[Perfect Recall]
    An agent $a \in \A$ has \textit{perfect recall} w.r.t. a total order $D_1,\ldots,D_n$ over $\D_a$ if for all $D_i,D_j \in \D_a$, $i<j$ implies that $D_i \in \pa(D_j)$ and $\pa(D_i) \subset \pa(D_j)$.
\end{definition}

\begin{proposition}[Correctness of Algorithm~\ref{alg:maid}~\cite{Koller2001}]
    Suppose every agent has perfect recall. Then, the output $\sigma^m$ of Algorithm~\ref{alg:maid} is a Nash equilibrium for $\M$.
\end{proposition}

\section{Model}\label{model}

Let us first describe a simple governance game. Assuming that there are $n$ homogeneous agents (meaning that all of these agents exhibit same exact behaviour given similar situations) who own stake in the governance of a DeFi protocol. Assume that the DeFi protocol has a governance token \$GOV and each of these agents owns some amount of \$GOV that would enable them to vote on proposals. Proposals can be brought forward by members of the governance for the rest to make a decision on. It is assumed that the member who brings in the proposal will choose to remain neutral and will abstain from voting. For example, in a DeFi lending borrowing protocol the proposal could be to strengthen the risk parameters of the collateral assets to allow a user to borrow less amount of an asset at the same amount of collateral simply to preserve the sustainability of the protocol. Let us visualise this model through the lens of MAIDs.  Here we define the input variables of the MAID $\M = (\A,\chi,\D,\U)$ in the following way: 
\begin{enumerate}
    \item $\A$: set of $n$ agents;
    \item $\D_a = \{d_a\}$: set of decision variables for each agent $a\in\A$, where $d_a = \{\texttt{yes},\texttt{no}\}$ (decisions about a proposal);
    \item $\U_a = \{u_a\}$: set of utility variables for each agent $a \in \A$ (How sustainable this decision would be for the protocol? Will it benefit the protocol?)
    \item $\chi = \{\texttt{CR},\texttt{MS}\}$: set of chance variables (collateral risk $\texttt{CR}$ and market sentiment $\texttt{MS}$ that determine the decision, where $\texttt{CR}\in \{\texttt{risky},\texttt{not risky}\}$ and $\texttt{MS}\in \{\texttt{good},\texttt{bad}\}$.
\end{enumerate}
We define utilities as follows: For agent $a \in \A$, 
\begin{align}
u_a(d_a|\texttt{CR}=\texttt{risky},\texttt{MS}=\texttt{bad})
        &=\begin{cases}
            +100& \text{ if } d_a = \texttt{yes},\\
            -100& \text{ if } d_a = \texttt{no}.
        \end{cases} \\
u_a(d_a|\texttt{CR}=\texttt{risky},\texttt{MS}=\texttt{good})
        &=\begin{cases}
            +50& \text{ if } d_a = \texttt{yes},\\
            -50& \text{ if } d_a = \texttt{no}.
        \end{cases} \\
u_a(d_a|\texttt{CR}=\texttt{not risky},\texttt{MS}=\texttt{bad})
        &=\begin{cases}
            +25& \text{ if } d_a = \texttt{yes},\\
            -25& \text{ if } d_a = \texttt{no}.
        \end{cases} \\
u_a(d_a|\texttt{CR}=\texttt{not risky},\texttt{MS}=\texttt{good})
        &=\begin{cases}
            -100& \text{ if } d_a = \texttt{yes},\\
            +100& \text{ if } d_a = \texttt{no}.
        \end{cases} 
\end{align}

\subsection{Explanation of the Utilities} 
As we have defined two chance variables---collateral risk and market sentiment---we need distinguish between four cases.
\begin{enumerate}
    \item For any agent $a \in \mathcal{A}$, given a collateral asset classified as risky and a prevailing negative market sentiment, voting in favor of the proposal (decision variable $d_a = \texttt{yes}$) yields a positive of $+100$. Conversely, voting against the proposal ($d_a = \texttt{no}$) results in a utility of $-100$.
    \item For any agent $a \in \mathcal{A}$, given a risky collateral asset and a positive market sentiment, voting in favor of the proposal (decision variable $d_a = \texttt{yes}$) results in a positive utility of $+50$. Conversely, voting against the proposal ($d_a = \texttt{no}$) leads to a utility of $-50$.

    \item For any agent $a \in \mathcal{A}$, given a non-risky collateral asset and a negative market sentiment, voting in favor of the proposal (decision variable $d_a = \texttt{yes}$) yields a positive utility of $+25$, primarily driven by anticipated future risk mitigation benefits. Analogously for $d_a = \texttt{no}$.

    \item For any agent $a \in \mathcal{A}$, given a non-risky collateral asset and a positive market sentiment, voting in favor of the proposal (decision variable $d_a = \texttt{yes}$) is deemed unsustainable and results in a utility decrement of 100. Analogously for $d_a = \texttt{no}$.

\end{enumerate}

\begin{remark}
We consider equal voting power regardless of the amount of \$GOV tokens that each agent has. Under stake-based voting, each agent's expected utility would be scaled by the proportion of \$GOV tokens.
In our model, we assume independence between the chance variables---market sentiment and collateral risk. We do not assume a direct causal link between market sentiment and collateral risk~\cite{pearl2009causality}.
We assume homogeneous agents to make the model tractable. 
\end{remark}

\section{Analysis} \label{analysis}
As devised in Section~\ref{MAIDs}, we present some experimental results using the "pycid" library developed by Google Deepmind~\cite{fox2021pycid}. We start by initialising and devising dummy variables as follows:
\begin{itemize}
    \item Recall collateral risk as $\texttt{CR}$. For $\texttt{CR}\in \{\texttt{risky},\texttt{not risky}\} $, we create dummy binary variables where $\texttt{risky}$ is denoted by $\texttt{1}$ and $\texttt{not risky}$ is denoted by $\texttt{0}$. When we say that collateral is "risky", we mean that the collateral is portraying high levels of price volatility at that instant of time and vice versa when we use the term "not risky".

    \item Recall market sentiment as $\texttt{MS}$. For $\texttt{MS}\in \{\texttt{good},\texttt{bad}\} $, we similarly create dummy binary variables where $\texttt{good}$ is denoted by $\texttt{0}$ and $\texttt{bad}$ is denoted by $\texttt{1}$. Again, when we say that market sentiment is "good", we mean that the market conditions for the collateral asset and future sentiment for its price volatility is favourable for long term investment and vice versa when we use the term "bad".

    \item Vote as $\texttt{V}$. Notice that $\texttt{V} = d_a$ and the notation was chosen for the slight difference to represent the action of voting and for ease of keeping track. Recall that  $\texttt{V} = d_a = \{\texttt{yes},\texttt{no}\} = \{\texttt{1},\texttt{0}\}$ for $a \in \A$. Simply, $\texttt{yes}$ implies voting for the proposal and $\texttt{no}$ is voting against the proposal. Every agent in the protocol has the same exact choices for the proposal. 

    \item Utility as $\texttt{U}$ as defined above as a utility function. 
\end{itemize}
\subsection{Single Agent Optimal Strategy} \label{single}

\begin{figure}[t]
    \includegraphics[width=0.5\linewidth]{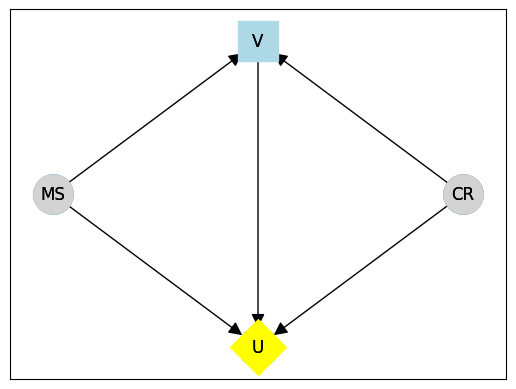}
    \caption{Causal Influence Diagram for Model}
    \label{fig1}
\end{figure}

We have all the tools that we need to analyse the model. Let us consider the case of a single agent in this  model of governance. Given inputs$ (\A,\chi,\D,\U)$ as we have defined above in Section~\ref{model} equations $(1),(2),(3)$ and $(4)$, we have all the ingredients required to develop the MAID for the single agent in question and
the dynamics of the governance protocol as defined above can be summarized in the causal influence diagram (CID) in Figure~\ref{fig1}.

$\texttt{CR}, \texttt{MS}$ influence the $\texttt{V}$ and $\texttt{U}$. We defined the conditional probability distribution (CPD) on the chance variables, in the form of a uniform distribution, as follows: 
$$P(\texttt{CR} = \texttt{risky(1)}) =  P(\texttt{CR} = \texttt{not risky(0)}) = 0.5$$
$$P(\texttt{MS} = \texttt{good(0)}) = P(\texttt{MS} = \texttt{bad(1)}) = 0.5$$

\begin{figure}[H]
    \includegraphics[width=0.5\linewidth]{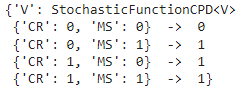}
    \caption{Optimal Policy for an Agent in the Game}
    \label{fig:nashh}
\end{figure}

Under the utility functions defined in the previous section and using the preliminaries, the optimal policy for $V = 0$ (Vote against) and $V = 1$ (Vote For) can be summarised in Figure~\ref{fig:nashh}.

Using definition of optimal strategy, for any agent $a \in \A$, 
$$EU_a(\sigma_V(1)) \geq EU_a(\sigma_V(0))$$ whenever $\texttt{CR} = 1$ or $\texttt{MS} = 1$. Similarly 
$$EU_a(\sigma_V(0)) \geq EU_a(\sigma_V(1))$$ when $\texttt{CR} = 0$ and $\texttt{MS} = 0$
That is, for any arbitrary agent in the MAID, the optimal strategy $\sigma$ for every agent would be to vote for strengthening the risk parameters $(\texttt{V} = 1)$ in case the collateral dynamics are risky $(\texttt{CR} = 1)$ or if the market sentiment is bad $(\texttt{MS} = 1)$and vote against the proposal $(\texttt{V} = 0)$ only if both the collateral dynamics are less volatile $(\texttt{CR} = 0)$ and the market sentiment is good $(\texttt{MS} = 0)$ too. Figure~\ref{fig:nashh} gives a good summary of this as an output using the pycid library. The expected utility for every single arbitrary agent is $\texttt{68.75}$, solved as per the preliminaries and the specifics of the model i.e. $EU(\sigma)  =\sum_{U \in \U_a} \sum_{u \in dom(U)}  P_{\M[\sigma]}(U = u) \cdot u =  {0.25} \cdot ({100 + 50 + 25 + 100 }){ = 68.75}$. Notice that in case the voting power was stake based, the expected utility would have been $\frac{\texttt{\$GOV owned by agent} }{\texttt{\$GOV total supply}} \cdot \texttt{68.75}$

\subsection{Multi Agent Equilibrium}\label{sec:multi}
To put things back into perspective, we know the optimal strategy for an agent in our  governance model and so we have an understanding of their independent decisions when faced with the same conditions. But what does it mean for their collective equilibrium? Let us formally address this. 

\begin{remark}
    If we assume that all agents in the governance protocol are making the decisions simultaneously, then it is safe to also assume that their independent decisions would converge to the same decision. So we make another assumption of non simultaneous decisions and every agent can see each other's vote to add in some complexity.
\end{remark}

\begin{remark}
    In a DeFi governance model where agents are homogeneous and share the same optimal strategy but make decisions at different times, voting on a proposal takes the form of a Stackelberg Game, where the leader acts according to their optimal policy first.
\end{remark}
To see this in action let us consider the case of two agents in the system and their interactions. Since the population of governance voters is assumed to be homogeneous and similar, taking case of two agents is sufficient. Like before let us define and initialise variables as follows:
\begin{itemize}
    \item Agent 1 as $\texttt{A1}$

    \item Agent 2 as $\texttt{A2}$

    \item Utility of Agent 1 as $\texttt{U1}$

    \item Utility of Agent 2 as $\texttt{U2}$
\end{itemize}

We change the perspective of the game. As we have assumed before, every agent $a \in \A$ has the same CID $(\V_a,E_a)$ which implies the same chance $\chi$, decision variables $\D_a$ and utility variables  $\U_a$ for all agents.  

\begin{figure}[t]
    \includegraphics[width=0.5\linewidth]{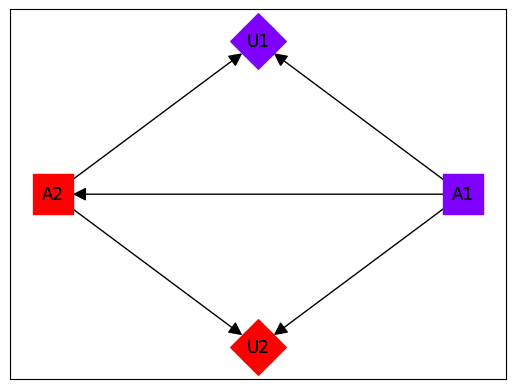}
    \caption{Multi Agent Causal Influence Diagram}
    \label{fig:MAID}
\end{figure}

Given some value of chance variables $\texttt{CR}$ and $\texttt{MS}$, agent $\texttt{A1}$ has made a decision to vote $\texttt{V}$ on the proposal to derive utility $\texttt{U1}$ which will be seen by the other agent in the governance protocol.  
The influence diagram (Figure~\ref{fig:MAID}) implies that each agent's decision has an effect on the other's utility. Recall the non simultaneous criterion that implies that one agent's vote influences the other agent's vote and hence also the decision. 
Consider time steps $\texttt{t} = \{1,2\}$ (i.e the game will run for $2$ time steps for each agent to make their decision). Both agents have the same chance variables and decision variables.
For $\texttt{t} = 1$, $\texttt{A1}$ makes a decision that gives him utility according to the scheme mentioned in Section~\ref{model}. 
At $\texttt{t} = 2$, $\texttt{A2}$ observes the decision made by $\texttt{A1}$ and makes their own decision. If agent $\texttt{A2}$ follows the decision of agent $\texttt{A1}$, it  also follows their optimal strategy. Hence, the utilities are equal, $\texttt{U1} = \texttt{U2}$. 

In all other cases we claim that the utilities would be $\leq 0$. Indeed, since agents $\texttt{A1}$ and $\texttt{A2}$ are following their optimal strategies and are homogeneous (similar), a positive utility $\texttt{U2}$ for $\texttt{A2}$ by choosing a different decision than $\texttt{A1}$ breaks the homogeneous assumption since that would imply that one of the two agents derive more utility from a decision than the other which is not possible.

\begin{remark}
    Given a system with agents \texttt{A1} and \texttt{A2}, where \texttt{A1} is designated as the leader, if agent \texttt{A2} experiences a net positive utility ($U_2 > 0$) by adopting a strategy divergent from \texttt{A1}'s, the following conditions must hold:
\begin{enumerate}
    \item Agent Heterogeneity: The utility functions of \texttt{A1} and \texttt{A2} are not identical, implying diverse preferences or objectives.
    \item Strategic Rationality: Agent \texttt{A2}'s decision to deviate is a strategic choice, prioritizing its individual utility maximization over collective welfare or protocol adherence.
\end{enumerate}

Under these conditions, \texttt{A2}'s behavior can be characterized as strategic or potentially adversarial, posing a risk to the system's equilibrium.
\end{remark}

\begin{figure}[t]
    \includegraphics[width=0.5\linewidth]{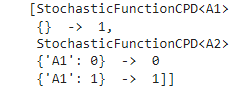}
    \caption{Pure Nash Equilibrium for Two Agents}
    \label{fig:pure}
\end{figure}

Using Algorithm~\ref{alg:maid}, we obtain the results for the pure Nash equilibrium of the system displayed in Figure~\ref{fig:pure}.
The analysis indicates that in a two-agent system where agent $\texttt{A1}$ acts as a leader, the optimal strategy for agent $\texttt{A2}$ is to mimic $\texttt{A1's}$ decision. This outcome aligns with the equilibrium of a Stackelberg game~\cite{julien2018stackelberg}, wherein the leader's choice precedes and influences the follower's action. Consequently, within a homogeneous agent population, where all agents possess identical utility functions and information, there exists a strong incentive for each agent to adopt the optimal policy of every other agent, leading to a collective action scenario. This observation supports the proposed hypothesis. 

What does it mean for DeFi governance protocols? One could claim that if the optimal strategy for any arbitrary agent in a DeFi governance is to promote protocol sustainability then the protocol should have no instances of manipulation (such as Governance Extractable Value~\cite{werner2022sok}) assuming homogeneity of agents. In case there is a deviation from the optimal strategy (or if there exists an incentive to deviate) then there will exist an instance of "attack" behavior which may be harmful to the protocol. In the next section, we look at a very simple attack model using MAIDs.

\subsection{A Simple Attack Model}
Previously, our model assumed homogeneous agents. We now introduce an adversarial agent to capture potential malicious behavior.
Formally, we extend the model in Section \ref{model} by introducing an additional agent $a' \in \mathcal{A}$ with identical chance variables $\chi$ and decision space $\mathcal{D}_{a'} = \mathcal{D}_a$ as the other agents. However, $a'$ possesses a distinct utility function $u_{a'}$. This adversarial agent aims to disrupt the protocol by acting contrary to the interests of honest agents.

Thus the utility function of agent $a'$ can be written out as follows:
\begin{align}
u_{a'}(d_{a'}|\texttt{CR}=\texttt{risky},\texttt{MS}=\texttt{bad})
        &=\begin{cases}
            -100& \text{ if } d_{a'} = \texttt{yes},\\
            +100& \text{ if } d_{a'} = \texttt{no}.
        \end{cases} \\
u_{a'}(d_{a'}|\texttt{CR}=\texttt{risky},\texttt{MS}=\texttt{good})
        &=\begin{cases}
            -50& \text{ if } d_{a'} = \texttt{yes},\\
            +50& \text{ if } d_{a'} = \texttt{no}.
        \end{cases} \\
u_{a'}(d_{a'}|\texttt{CR}=\texttt{not risky},\texttt{MS}=\texttt{bad})
        &=\begin{cases}
            -25& \text{ if } d_{a'} = \texttt{yes},\\
            +25& \text{ if } d_{a'} = \texttt{no}.
        \end{cases} \\
u_{a'}(d_{a'}|\texttt{CR}=\texttt{not risky},\texttt{MS}=\texttt{good})
        &=\begin{cases}
            +100& \text{ if } d_{a'}= \texttt{yes},\\
            -100& \text{ if } d_{a'} = \texttt{no}.
        \end{cases} 
\end{align}
Consider a governance proposal to strengthen risk parameters for a risky collateral asset. Under standard assumptions of agent homogeneity, the optimal strategy and Nash equilibrium for agents is to vote "yes" ($d_a = \texttt{yes}$), yielding a utility of $+100$.

We introduce an adversarial agent $a' \in \mathcal{A}$ with the primary objective of disrupting the protocol. In contrast to honest agents, $a'$ optimally votes "no" ($d_{a'} = \texttt{no}$), resulting in a utility of $+100$. 

\begin{figure}[t]
    \includegraphics[width=0.5\linewidth]{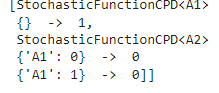}
    \caption{Nash Equilibrium for Honest Agent vs Attacker }
    \label{fig:attack}
\end{figure}

Formally, the system comprises two agents, $a$ and $a'$, representing honest and adversarial actors, respectively. Assuming identical CID and MAID structures as in Section \ref{sec:multi}, the system's architecture aligns with Figure \ref{fig:MAID} where $A_2 = a'$ and $A_1 = a$.

where $\texttt{U1}$ and $\texttt{U2}$ are the utilities of the honest agent $a$ and attacker $a'$ respectively.
We construct a payoff matrix for both the agents through their utilities given a decision that they make about the proposal given above. The payoff matrix can be summarised as follows:
$$
\begin{tabular}{|c|c|c|} \hline  
a/a'& 0 & 1 \\ \hline  
0  & 0,100& -100,-100 \\ \hline  
1  & 100,100& 100,-100\\ \hline 
\end{tabular}
$$
The Nash equilibrium for this set-up using Algorithm~\ref{alg:maid} is given in Figure~\ref{fig:attack}.

Solving the MAID reveals that agent $a'$ adopts a dominant strategy of choosing $\texttt{0}$, irrespective of the decision of $a$. Conversely, $a$ exhibits a dominant strategy of selecting $\texttt{1}$, independent of the choice of $a'$. Both agents have an expected utility of $\texttt{0}$. Honest agents achieve non-negative net utility when constituting a majority, whereas adversarial agents attain non-negative utility under majority control.
The application of MAIDs effectively demonstrates the framework's capacity to formally verify agent incentives within a governance protocol, highlighting its accessibility and utility for analyzing strategic interactions.

\section{Application to MakerDAO Governance Protocol}\label{sec:maker}
The Maker Protocol operates as a decentralized finance (DeFi) platform on the Ethereum blockchain, facilitating the creation of Dai, a stablecoin pegged to the US dollar. As per~\cite{maker}, Users engage with the protocol by depositing cryptocurrency assets, such as Ethereum or other tokens, into smart contracts as collateral. Based on the value of this collateral, users can then generate Dai, typically up to a predetermined percentage of the collateral's worth. To maintain the stability and security of the system, the protocol enforces various risk parameters, including collateralization ratios and debt ceilings, which dictate the relationship between collateral and the amount of Dai that can be generated. Users pay stability fees on the Dai they generate, serving as interest rates that contribute to the protocol's stability. Governance of the Maker Protocol is decentralized, with MKR token holders participating in decision-making through a voting mechanism. These holders vote on proposals to adjust risk parameters and other protocol settings, ensuring the protocol remains responsive to market conditions and emerging risks. Additionally, the protocol includes mechanisms for liquidating collateral if its value falls below a certain threshold, helping to mitigate systemic risk and maintain the stability of the Dai ecosystem. More information about the governance dynamics of the Maker protocol can be found in~\cite{SOKG}.

The binary voting mechanism employed by MakerDAO, where MKR holders cast yes/no votes on proposals, is analogous to the governance games analyzed in preceding sections. By applying MAIDs to a simplified binary voting model, we can examine agent interactions under two scenarios: homogeneous agents sharing identical MAIDs and heterogeneous agents with opposing objectives (e.g., protocol sustainability vs. disruption). This analysis provides insights into the game-theoretic dynamics within the protocol and MAIDs can be used in similar ways to model more complex interactions and in turn be used to create a formal verification system for coordination games in DeFi governance protocols. 

\subsection{MAID for MakerDAO}
\begin{figure*}[t]
    \centering
    \includegraphics[width=0.7\textwidth]{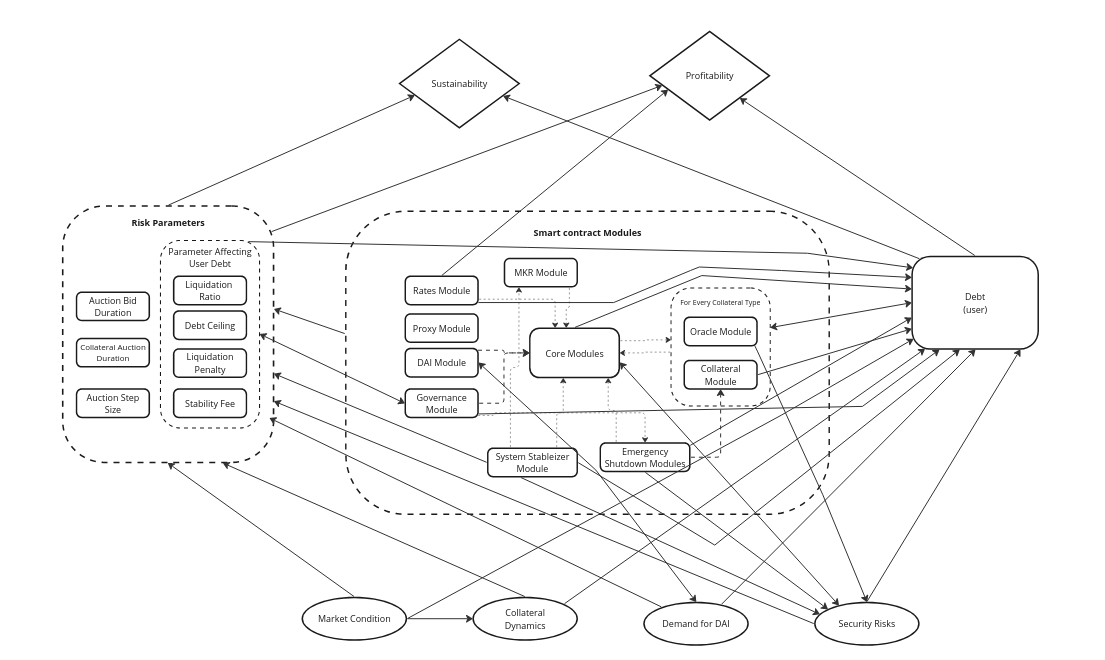}
    \caption{MAID for Maker on the Smart Contract Level}
    \label{MakerMAID}
\end{figure*}

In this section, we outline the structure of a game between the users of the Maker protocol and the MakerDAO governance member on the smart contract level.
For the MAID, the variables are as follows:
\begin{itemize}
    \item The chance variables $\chi$ include smart contract modules, market conditions, collateral dynamics, demand for DAI stablecoin and security risks.
    \item The decision variables $\D$ include debt (that the user wants to undertake) and the risk parameters (that the governance DAO members want to set for a particular asset).
    \item The utility variables $\U$ include profitability for the users of the protocol and protocol sustainability for the governance members of the DAO.
\end{itemize}
The MAID is depicted in Figure~\ref{MakerMAID}.

\subsubsection{Chance Variables}
First, we outline the various smart contracts modules~\cite{maker2014} that interact with each other and with external actors to maintain system stability and facilitate user interactions. Then, we describe the remaining chance variables.
\begin{enumerate}
    \item {Smart Contract Modules}:
\begin{itemize}
    \item {\textit{Core Module}}: The Core Module is crucial to the system as it contains the entire state of the Maker Protocol and controls the central mechanisms of the system while it is in the expected normal state of operation.
    \item {\textit{Collateral Module}}: The collateral module is deployed for every new \verb|ilk| (collateral type) added to \verb|Vat|. It contains all the adapters and auction contracts for one specific collateral type.
    \item {\textit{The System Stabilizer Module}}: It's purpose is to correct the system when the value of the collateral backing Dai drops below the liquidation level (determined by governance) when the stability of the system is at risk. The system stabilizer module creates incentives for Auction Keepers (external actors) to step in and drive the system back to a safe state (system balance) by participating in both debt and surplus auctions and, in turn, earn profits by doing so.
    \item {\textit{Oracle Module}}: An oracle module is deployed for each collateral type, feeding it the price data for a corresponding collateral type to the \verb|Vat|. The Oracle Module introduces the whitelisting of addresses, which allows them to broadcast price updates off-chain, which are then fed into a \verb|median| before being pulled into the \verb|OSM|. The \verb|Spot|'ter will then proceed to read from the \verb|OSM| and will act as the liaison between the \verb|oracles| and \verb|dss|.
    \item {\textit{MKR Module}}: The MKR Module contains the MKR token, which is a deployed DS Token contract. It is an ERC20 token that provides a standard ERC20 token interface. It also contains logic for burning and authorized minting of MKR.
    \item {\textit{Governance Module}}: The Governance Module contains the contracts that facilitate MKR voting, proposal execution, and voting security of the Maker Protocol.The Governance Module has 3 core components consisting of the \verb|Chief|, \verb|Pause| and \verb|Spell| contracts.
    \item {\textit{Rate Module}}: The Maker Protocol's Rate Accumulation Mechanism
    \item {\textit{Proxy Module}}: The Proxy module was created in order to make it more convenient for users/developers to interact with the Maker Protocol. It contains contract interfaces, proxies, and aliases to functions necessary for both DSR and Vault management and Maker governance.
    \item {\textit{Maker Protocol Emergency Shutdown:}} Shutdown is a process that can be used as a last resort to directly enforce the Target Price to holders of Dai and Vaults, and protect the Maker Protocol against attacks on its infrastructure. Shutdown stops and gracefully settles the Maker Protocol while ensuring that all users, both Dai holders and Vault holders, receive the net value of assets they are entitled to.
\end{itemize}

\item \textit{Market Conditions}: Market conditions significantly impact the stability and performance of the Maker Protocol. Risk parameters such as the liquidation ratio and penalty are designed to account for market fluctuations and maintain system health during adverse conditions. The Oracle Module provides real-time price data, enabling swift reactions to changing market conditions, while dynamically adjusting debt ceilings based on market conditions helps manage risk and maintain stability.

\item \textit{Collateral Dynamics}: The Collateral Module manages collateral types and associated parameters within the Maker Protocol. Adjusting risk parameters like the debt ceiling and liquidation ratio influences collateral dynamics by affecting the selection and utilization of collateral assets. Furthermore, the efficiency of collateral auctions, facilitated by modules like the System Stabilizer Module, impacts collateral dynamics by determining the ease of collateral liquidation in case of undercollateralization.

\item \textit{Demand for DAI}: The demand for DAI is intricately linked to borrowing conditions within the Maker Protocol, influenced primarily by the stability fee. Lower stability fees make borrowing DAI more attractive, potentially increasing demand, although this must be balanced with the need for protocol revenue. Additionally, risk parameters such as the liquidation ratio and debt ceiling play crucial roles in managing risk and ensuring the stability of DAI, thus impacting demand. The efficiency of collateral auctions facilitated by modules like the System Stabilizer Module also influences confidence in the protocol and, consequently, demand for DAI.

\item \textit{Security Risks}: Security risks within the Maker Protocol stem from vulnerabilities in smart contracts, market manipulation, or unexpected events like oracle failures. The Oracle Module is vital for mitigating security risks by providing accurate price feeds, while the Maker Protocol Emergency Shutdown mechanism serves as a safeguard against critical failures. Adjusting parameters like the liquidation ratio and debt ceiling is essential for managing security risks associated with insolvency or exposure to volatile assets.

\end{enumerate}

\subsubsection{Decision Variables}
\begin{enumerate}
    \item {\textit{Debt:}}
The debt within the Maker Protocol represents borrowed Dai against provided collateral. Smart contract modules like the Core Module and Rate Module are instrumental in managing user debt by maintaining accurate records and dynamically adjusting stability fees based on market conditions. Risk parameters such as the debt ceiling, liquidation ratio, and penalty further influence debt dynamics by setting limits and incentivizing adequate collateralization.
    \item {\textit{Risk Parameters:}} Risk parameters, governed by the Governance Module, play a pivotal role in capital efficiency, risk mitigation, and profitability within the Maker Protocol ecosystem. Parameters such as the liquidation ratio, debt ceiling, liquidity penalty, and stability fee directly influence user debt, security risk, demand for DAI, and collateral dynamics. These parameters must be meticulously adjusted to ensure high capital efficiency, low risk exposure, and sustainable profitability. For instance, the debt ceiling sets limits on debt generation against specific collateral types to prevent over-leveraging and maintain system stability. Similarly, the liquidation ratio and penalty incentivize users to uphold sufficient collateralization levels, mitigating the risk of liquidation events.

\end{enumerate}

\subsubsection{Utility Variables}
\begin{enumerate}
    \item \textit{Profitability:} Profitability within the Maker Protocol ecosystem is intricately linked to the effectiveness of risk parameters and smart contract modules. The stability fee, determined by governance, directly impacts the profitability of Vault owners by influencing the cost of borrowing Dai against collateral. Additionally, the efficacy of the System Stabilizer Module in managing liquidations and surplus auctions affects the profitability of participants, including Keepers engaging in these auctions to earn profits. Higher liquidation penalties incentivize users to maintain adequate collateralization, thereby reducing the need for liquidations and potential losses for the protocol.
    
    \item \textit{Sustainability:} The sustainability of the Maker Protocol hinges on prudent management of risk parameters such as the debt ceiling, liquidation ratio, and stability fee. While higher debt ceilings can potentially increase revenue through stability fees, they also expose the protocol to greater risk if not managed carefully. Similarly, a lower liquidation ratio allows for more borrowing and higher fee generation but escalates the risk of undercollateralization and losses during market downturns. The stability fee, crucial for revenue generation, must strike a delicate balance to avoid discouraging borrowing and reducing demand for DAI. The Governance Module assumes a pivotal role in aligning decisions regarding risk parameters and protocol upgrades with sustainability goals.
\end{enumerate}

\section{Conclusion and Future Work}\label{sec:conclusion}
In this study, we explored a simple DeFi governance game and analyzed its equilibria in both single and multi-agent settings using Multi-Agent Influence Diagrams (MAIDs). Our analysis included a practical example involving an attacker and provided the framework for constructing a MAID specifically tailored to MakerDAO, which can serve as a foundation for further exploration of decentralized governance dynamics.
Our work demonstrates the potential of MAIDs as a powerful tool for understanding the complex interactions within DeFi governance models. However, there are several promising directions for future research. Firstly, our current model employs discrete uniform conditional probability distributions; extending this to continuous distributions could yield more insights. Additionally, the analysis was limited to two agents, and there is significant potential to expand this framework to accommodate a larger number of agents, each with possibly different utility functions, to better mirror the diverse landscape of DeFi ecosystems.
Moreover, while our study does not assume a direct causal link between market sentiment and collateral risk, incorporating such causation could provide a deeper understanding of the effects on governance outcomes.
Solving the MAID for MakerDAO using real-world parameters remains an open challenge and will be crucial for validating the applicability of our model. .
As DeFi continues to grow, such tools will be instrumental in guiding the development of secure and resilient governance structures.

\bibliographystyle{ACM-Reference-Format}
\bibliography{references.bib}


\begin{thebibliography}{15}


\ifx \showCODEN    \undefined \def \showCODEN     #1{\unskip}     \fi
\ifx \showDOI      \undefined \def \showDOI       #1{#1}\fi
\ifx \showISBNx    \undefined \def \showISBNx     #1{\unskip}     \fi
\ifx \showISBNxiii \undefined \def \showISBNxiii  #1{\unskip}     \fi
\ifx \showISSN     \undefined \def \showISSN      #1{\unskip}     \fi
\ifx \showLCCN     \undefined \def \showLCCN      #1{\unskip}     \fi
\ifx \shownote     \undefined \def \shownote      #1{#1}          \fi
\ifx \showarticletitle \undefined \def \showarticletitle #1{#1}   \fi
\ifx \showURL      \undefined \def \showURL       {\relax}        \fi
\providecommand\bibfield[2]{#2}
\providecommand\bibinfo[2]{#2}
\providecommand\natexlab[1]{#1}
\providecommand\showeprint[2][]{arXiv:#2}

\bibitem[Cormen et~al\mbox{.}(2022)]%
        {cormen2022introduction}
\bibfield{author}{\bibinfo{person}{Thomas~H Cormen}, \bibinfo{person}{Charles~E Leiserson}, \bibinfo{person}{Ronald~L Rivest}, {and} \bibinfo{person}{Clifford Stein}.} \bibinfo{year}{2022}\natexlab{}.
\newblock \bibinfo{booktitle}{\emph{Introduction to algorithms}}.
\newblock \bibinfo{publisher}{MIT press}.
\newblock


\bibitem[Everitt et~al\mbox{.}(2021)]%
        {everitt2021agent}
\bibfield{author}{\bibinfo{person}{Tom Everitt}, \bibinfo{person}{Ryan Carey}, \bibinfo{person}{Eric~D Langlois}, \bibinfo{person}{Pedro~A Ortega}, {and} \bibinfo{person}{Shane Legg}.} \bibinfo{year}{2021}\natexlab{}.
\newblock \showarticletitle{Agent incentives: A causal perspective}. In \bibinfo{booktitle}{\emph{Proceedings of the AAAI Conference on Artificial Intelligence}}, Vol.~\bibinfo{volume}{35}. \bibinfo{pages}{11487--11495}.
\newblock


\bibitem[Everitt et~al\mbox{.}(2019)]%
        {everitt2019understanding}
\bibfield{author}{\bibinfo{person}{Tom Everitt}, \bibinfo{person}{Pedro~A Ortega}, \bibinfo{person}{Elizabeth Barnes}, {and} \bibinfo{person}{Shane Legg}.} \bibinfo{year}{2019}\natexlab{}.
\newblock \showarticletitle{Understanding agent incentives using causal influence diagrams. Part I: Single action settings}.
\newblock \bibinfo{journal}{\emph{arXiv preprint arXiv:1902.09980}} (\bibinfo{year}{2019}).
\newblock


\bibitem[Foundation(2014)]%
        {maker2014}
\bibfield{author}{\bibinfo{person}{Maker Foundation}.} \bibinfo{year}{2014}\natexlab{}.
\newblock \bibinfo{title}{\textit{Maker Smart Contract Module}}.
\newblock
\newblock
\urldef\tempurl%
\url{https://docs.makerdao.com/}
\showURL{%
\tempurl}


\bibitem[Fox et~al\mbox{.}(2021)]%
        {fox2021pycid}
\bibfield{author}{\bibinfo{person}{James Fox}, \bibinfo{person}{Tom Everitt}, \bibinfo{person}{Ryan Carey}, \bibinfo{person}{Eric~D Langlois}, \bibinfo{person}{Alessandro Abate}, {and} \bibinfo{person}{Michael~J Wooldridge}.} \bibinfo{year}{2021}\natexlab{}.
\newblock \showarticletitle{PyCID: A Python Library for Causal Influence Diagrams.}. In \bibinfo{booktitle}{\emph{SciPy}}. \bibinfo{pages}{65--73}.
\newblock


\bibitem[Howard and Matheson(2005)]%
        {howard2005influence}
\bibfield{author}{\bibinfo{person}{Ronald~A Howard} {and} \bibinfo{person}{James~E Matheson}.} \bibinfo{year}{2005}\natexlab{}.
\newblock \showarticletitle{Influence diagrams}.
\newblock \bibinfo{journal}{\emph{Decision Analysis}} \bibinfo{volume}{2}, \bibinfo{number}{3} (\bibinfo{year}{2005}), \bibinfo{pages}{127--143}.
\newblock


\bibitem[Julien(2018)]%
        {julien2018stackelberg}
\bibfield{author}{\bibinfo{person}{Ludovic~A Julien}.} \bibinfo{year}{2018}\natexlab{}.
\newblock \showarticletitle{Stackelberg games}.
\newblock In \bibinfo{booktitle}{\emph{Handbook of Game Theory and Industrial Organization, Volume I}}. \bibinfo{publisher}{Edward Elgar Publishing}, \bibinfo{pages}{261--311}.
\newblock


\bibitem[Kiayias and Lazos(2022)]%
        {SOKG}
\bibfield{author}{\bibinfo{person}{Aggelos Kiayias} {and} \bibinfo{person}{Philip Lazos}.} \bibinfo{year}{2022}\natexlab{}.
\newblock \showarticletitle{SoK: blockchain governance}. In \bibinfo{booktitle}{\emph{Proceedings of the 4th ACM Conference on Advances in Financial Technologies}}. \bibinfo{pages}{61--73}.
\newblock


\bibitem[Koller and Milch(2001)]%
        {Koller2001}
\bibfield{author}{\bibinfo{person}{Daphne Koller} {and} \bibinfo{person}{Brian Milch}.} \bibinfo{year}{2001}\natexlab{}.
\newblock \showarticletitle{Multi-agent influence diagrams for representing and solving games}. In \bibinfo{booktitle}{\emph{Proceedings of the 17th International Joint Conference on Artificial Intelligence - Volume 2}} (Seattle, WA, USA) \emph{(\bibinfo{series}{IJCAI'01})}. \bibinfo{publisher}{Morgan Kaufmann Publishers Inc.}, \bibinfo{address}{San Francisco, CA, USA}, \bibinfo{pages}{1027–1034}.
\newblock
\showISBNx{1558608125}


\bibitem[MakerDAO(2024)]%
        {maker}
\bibfield{author}{\bibinfo{person}{MakerDAO}.} \bibinfo{year}{2024}\natexlab{}.
\newblock \bibinfo{title}{MakerDAO Voting Portal}.
\newblock \bibinfo{howpublished}{\url{https://vote.makerdao.com/}}.
\newblock
\newblock
\shownote{Accessed: 2024-07-23}.


\bibitem[Nash et~al\mbox{.}(1950)]%
        {nash1950non}
\bibfield{author}{\bibinfo{person}{John~F Nash} {et~al\mbox{.}}} \bibinfo{year}{1950}\natexlab{}.
\newblock \showarticletitle{Non-cooperative games}.
\newblock  (\bibinfo{year}{1950}).
\newblock


\bibitem[Pearl(2009)]%
        {pearl2009causality}
\bibfield{author}{\bibinfo{person}{Judea Pearl}.} \bibinfo{year}{2009}\natexlab{}.
\newblock \bibinfo{booktitle}{\emph{Causality}}.
\newblock \bibinfo{publisher}{Cambridge university press}.
\newblock


\bibitem[Pearl(2014)]%
        {pearl2014probabilistic}
\bibfield{author}{\bibinfo{person}{Judea Pearl}.} \bibinfo{year}{2014}\natexlab{}.
\newblock \bibinfo{booktitle}{\emph{Probabilistic reasoning in intelligent systems: networks of plausible inference}}.
\newblock \bibinfo{publisher}{Elsevier}.
\newblock


\bibitem[Werner et~al\mbox{.}(2022)]%
        {werner2022sok}
\bibfield{author}{\bibinfo{person}{Sam Werner}, \bibinfo{person}{Daniel Perez}, \bibinfo{person}{Lewis Gudgeon}, \bibinfo{person}{Ariah Klages-Mundt}, \bibinfo{person}{Dominik Harz}, {and} \bibinfo{person}{William Knottenbelt}.} \bibinfo{year}{2022}\natexlab{}.
\newblock \showarticletitle{Sok: Decentralized finance (defi)}. In \bibinfo{booktitle}{\emph{Proceedings of the 4th ACM Conference on Advances in Financial Technologies}}. \bibinfo{pages}{30--46}.
\newblock


\bibitem[Williams et~al\mbox{.}(2018)]%
        {williams2018directed}
\bibfield{author}{\bibinfo{person}{Thomas~C Williams}, \bibinfo{person}{Cathrine~C Bach}, \bibinfo{person}{Niels~B Matthiesen}, \bibinfo{person}{Tine~B Henriksen}, {and} \bibinfo{person}{Luigi Gagliardi}.} \bibinfo{year}{2018}\natexlab{}.
\newblock \showarticletitle{Directed acyclic graphs: a tool for causal studies in paediatrics}.
\newblock \bibinfo{journal}{\emph{Pediatric research}} \bibinfo{volume}{84}, \bibinfo{number}{4} (\bibinfo{year}{2018}), \bibinfo{pages}{487--493}.
\newblock


\end{thebibliography}

\end{document}